\documentclass[12pt,draftcls,onecolumn]{IEEEtran}
\usepackage{cite}
\usepackage[font=small]{caption}
\usepackage{ifpdf}
\usepackage{authblk}
\usepackage{commath}
\usepackage{cuted} 

\usepackage{bigints}
\newif\ifCLASSOPTIONonecolumn       \CLASSOPTIONonecolumnfalse
\newif\ifCLASSOPTIONtwocolumn       \CLASSOPTIONtwocolumntrue
\ifCLASSINFOpdf
   \usepackage[pdftex]{graphicx}
 \else
    \usepackage[dvips]{graphicx}
  \fi
\usepackage{subcaption}
\usepackage{amsmath}
\usepackage{dsfont}
\usepackage{bbold}
\usepackage{amssymb}
\usepackage{algorithmic}
\usepackage{array}
\usepackage{stix}
\usepackage{stfloats}
\fnbelowfloat
\usepackage{bigints}
\begin{document}
%
\title{Non-Orthogonal Multiple Access Performance for Millimeter Wave in Vehicular Communications
}%
%
%
\author[1 2]{Baha Eddine Youcef~Belmekki}
\author[1]{Abdelkrim ~Hamza}
\author[2]{Beno\^it~Escrig}
\affil[1]{LISIC Laboratory, Electronic and Computer Faculty, USTHB, Algiers, Algeria,}
\affil[ ]{email: $\{$bbelmekki, ahamza$\}$@usthb.dz}
\affil[2]{University of Toulouse, IRIT Laboratory, School
of ENSEEIHT, Institut National Polytechnique de Toulouse, France, e-mail: $\{$bahaeddine.belmekki, benoit.escrig$\}$@enseeiht.fr}
\affil[ ]{}
\setcounter{Maxaffil}{0}
\renewcommand\Affilfont{\small}
\markboth{ }%
{Shell \MakeLowercase{\textit{et al.}}: Bare Demo of IEEEtran.cls for IEEE Journals}
\maketitle

\IEEEpeerreviewmaketitle
\begin{abstract}
 In this paper, we study the performance of millimeter wave (mmWave) vehicular communications (VCs) using non-orthogonal multiple access scheme (NOMA) at road intersections, since there areas are more prone to accidents. We study the case when the intersection involves two perpendicular lanes, we then extend the study to an intersection with several lanes. The transmission occurs between a source, and two destinations.
 The transmission experiences interference originated from a set of vehicles that are distributed as a Poisson point process (PPP).
 Our analysis includes the effects of blockage from buildings and vehicles at intersections. 
 Closed form outage probability expressions are obtained.
 We show that as the nodes reach the intersection, the outage probability increases.
 Counter-intuitively, we show that the non line of sigh (NLOS) scenario has a better performance than the line of sigh (LOS) scenario. 
 Finally, we compare the performance of mmWave NOMA with OMA, and we show that NOMA offers a significant improvement over OMA mmWave vehicular networks.
 The analysis is verified by Monte-Carlo simulation.
\end{abstract}

\section{Introduction}

\subsection{Motivation}
Road traffic accidents are a major issue, especially at road intersections \cite{traficsafety}.
In that regard, vehicular communications (VCs) offer several applications for road safety and traffic management. 
These applications can prevent accidents or alerting vehicles of accidents happening in their vicinity.
Hence, these applications need high data rate and high spectral efficiency, to insure high reliability and low latency transmissions.
In this context, non-orthogonal multiple access (NOMA) has been show to increase the data rate and spectral efficiency \cite{ding2017application}. Unlike orthogonal multiple access (OMA), NOMA allows multiple users to share the same resource with different power allocation levels.
On the other hand, the needs of VCs for the fifth generation (5G) of wireless networks in terms of resources require a larger bandwidth. Since the spectral efficiency of sub-6 GHz bands has already reached the theoretical limits, millimeter wave (mmWave) frequency bands (20-100 GHz and beyond) offer a very large bandwidth \cite{roh2014millimeter}.

\subsection{Related Works}
\subsubsection{NOMA Works}
NOMA is an efficient multiple access technique for spectrum use. It has been shown that NOMA outperforms OMA (see \cite{mobini2017full} and reference therein). However, few works investigate the effect of co-channel interference and their impact on the performance \cite{ali2018analyzing,zhang2016stochastic,tabassum2017modeling}.
The authors in \cite{ali2018analyzing} and \cite{zhang2016stochastic} analyze downlinks of NOMA networks. The authors in \cite{zhang2017uplink} and \cite{tabassum2017modeling} analyze uplinks of NOMA networks. In \cite{ali2017non}, the authors analyze the performance of NOMA transmissions and propose an interference aware NOMA design that takes into account both intercell and intracell interference.

\subsubsection{mmWave Works}
In mmWave bands, few works studied communications using tools from stochastic geometry \cite{biswas2016performance,wu2017coverage,belbase2018two,belbase2018coverage}. However, in \cite{biswas2016performance,wu2017coverage,belbase2018two}, the effect of small-scale fading is not taken into consideration. In \cite{belbase2018coverage}, the authors investigate the performance of mmWave relaying networks in terms of coverage probability with best relay selection.

\subsubsection{VCs at Road Intersections Works}
Several works studied the effect of the interference at intersections, considering OMA. The performance in terms of success probability are derivated considering direct transmission in \cite{steinmetz2015stochastic,abdulla2016vehicle}. The performance of vehicle to vehicle (V2V) communications are evaluated for multiple intersections schemes considering direct transmission in \cite{jeyaraj2017reliability}. In \cite{kimura2017theoretical}, the authors derive the outage probability of a V2V communications with power control strategy of a direct transmission. 
In \cite{article}, the authors investigate the impact of a line of sight and non line of sight transmissions at intersections considering Nakagami-$m$ fading channels. The authors in \cite{belmekki2018performance} study the effect of mobility of vehicular communications at road junctions.
In \cite{WCNC,VTC,WiMob,J3,J4}, the authors respectively study the impact of non-orthogonal multiple access, cooperative non-orthogonal multiple access, and maximum ratio combining with NOMA at intersections.
Following this line of research, we study the performance of VCs at intersections in the presence of interference. 
However, at the best of the author's knowledge, there are no prior works that consider both an intersection scenario with NOMA and considering mmWave networks. Our analysis includes the effects of blockage from the building and vehicles at intersections, and Nakagami-$m$ fading channels between the transmitting nodes with difference values of $m$ for LOS and NLOS are considered. Unlike other works that uses approximations, closed form expressions are obtained for Nakagami-$m$ fading channels.

\section{System Model}

\begin{figure}[]
\centering
\includegraphics[scale=0.5]{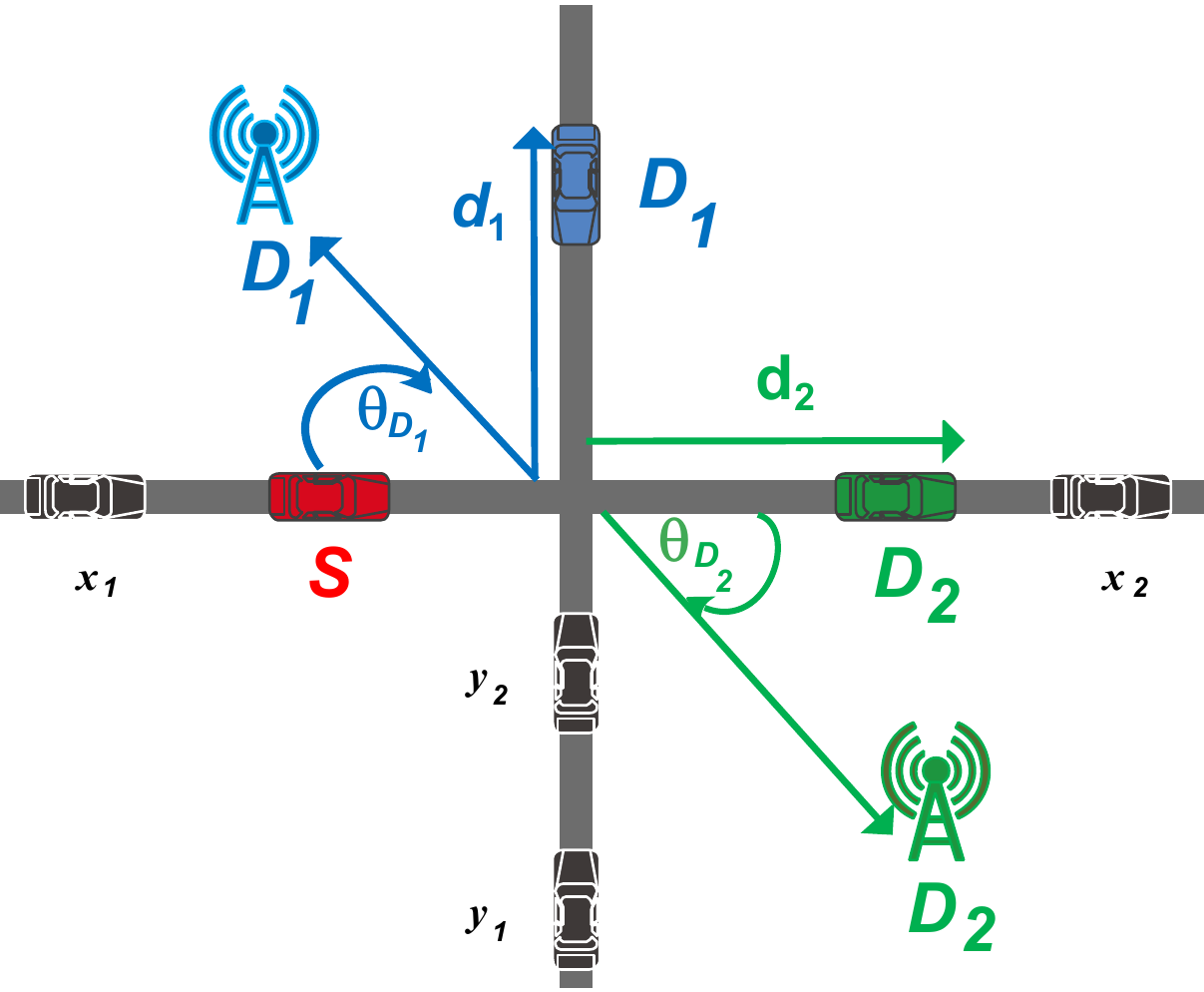}
\caption{NOMA system model for mmWave VCs involving a source and two destinations. The nodes can be vehicles or as part of the communication infrastructure.}
\label{Fig0}
\end{figure}

\subsection{Scenario Model}
We consider a mm-Wave vehicular network using a NOMA transmission between a source $S$, and two destinations $D_1$ and $D_2$. We consider an intersection scenario with two perpendicular roads, an horizontal road $X$, and a vertical road $Y$.

Since both V2V and V2I communications\footnote{The Doppler shift and time-varying effect of V2V and V2I channels are beyond the scope of this paper.} are considered, any node of the triplet $\lbrace{S, D_1, D_2}\rbrace$ can be on the road (e.g., vehicle), or outside the roads (e.g., infrastructure). We denote by  $D_i$ the receiving node, and by $d_i$ the distance between the node $D_i$ and the intersection, where $i \in \{1,2\}$, as shown in Fig.\ref{Fig0}. The angle $\theta_{D_i} $ is the angle between the node ${D_i}$ and the $X$ road (see Fig.\ref{Fig0}).

The transmission is subject to interference that are originated from vehicles located on the roads. The set of interfering vehicles located on the $X$ road that are in a LOS with $\lbrace{S, D_1, D_2}\rbrace$, denoted by $\Phi^{\textrm{LOS}}_{X}$ (resp. on the $Y$ road, denoted by $\Phi^{\textrm{LOS}}_{Y}$) are modeled as a one-dimensional homogeneous Poisson point process (1D-HPPP), that is, $\Phi^{\textrm{LOS}}_{X}\sim\textrm{1D-HPPP}(\lambda^{\textrm{LOS}}_{X},x)$ (resp.$\Phi^{\textrm{LOS}}_{Y}$ $\sim \textrm{1D-HPPP}(\lambda^{\textrm{LOS}}_{Y},y)$, where $x$ and $\lambda^{\textrm{LOS}}_{X}$ (resp. $y$ and $\lambda^{\textrm{LOS}}_{Y}$) are the position of the LOS interfering vehicles and their intensity on the $X$ road (resp. $Y$ road). 

Similarly, the set of interfering vehicles located on the $X$ road that are in a NLOS with $\lbrace{S, D_1, D_2}\rbrace$, denoted by $\Phi^{\textrm{NLOS}}_{X}$ (resp. on the $Y$ road, denoted by $\Phi^{\textrm{NLOS}}_{Y}$) are modeled as a 1D-HPPP, that is, $\Phi^{\textrm{NLOS}}_{X}\sim\textrm{1D-HPPP}(\lambda^{\textrm{NLOS}}_{X},x)$ (resp.$\Phi^{\textrm{NLOS}}_{Y}$ $\sim \textrm{1D-HPPP}(\lambda^{\textrm{NLOS}}_{Y},y)$, where $x$ and $\lambda^{\textrm{NLOS}}_{X}$ (resp. $y$ and $\lambda^{\textrm{NLOS}}_{Y}$) are the position of the NLOS interfering vehicles and their intensity on the $X$ road (resp. $Y$ road). The notation $x$ and $y$ denotes both the interfering vehicles and their locations.

\subsection{Blockage Model}
At the intersection, the mmWave signals cannot penetrate the obstacles (e.g., building, vehicles), which causes the link to be in LOS, or in NLOS. 
The event of a link between a node $a$ and $b$ is in a LOS and NLOS, are respectively defined as $\textrm{LOS}_{ab}$, and $\textrm{NLOS}_{ab}$.
The LOS probability function $\mathbb{P}(\textrm{LOS}_{ab})$ is used, where the link between $a$ and $b$
has a LOS probability $\mathbb{P}(\textrm{LOS}_{ab}) = \exp(-\beta r_{ab} )$ and NLOS probability
$\mathbb{P}(\textrm{NLOS}_{ab}) = 1-\mathbb{P}(\textrm{LOS}_{ab})$, where the constant rate $\beta$
depends on the building size, shape and density \cite{bai2014analysis}.

\subsection{Transmission Model}
The transmission between the nodes $a$ and $b$ experiences a path loss, denoted $r_{ab}^{-\alpha}$ , where $ r_{ab}=\Vert a- b\Vert$, and $\alpha$ is the path loss exponent. The path exponent $\alpha \in \{\alpha_{\textrm{LOS}}, \alpha_{\textrm{NLOS}}\}$, where $\alpha=\alpha_{\textrm{LOS}}$, when the transmission is in LOS, whereas $\alpha=\alpha_{\textrm{NLOS}}$, when transmission is in NLOS.

\subsection{Medium Access Control (MAC) Protocol}

The medium access protocols used in VCs are mainly based on carrier sense multiple access (CSMA) schemes (e.g., IEEE 802.11 p) \cite{karagiannis2011vehicular}. However, \cite{subramanian2012congestion,nguyen2013performance} showed that the performance of CSMA tends to the performance of ALOHA in dense networks. Hence, we assume that vehicles use slotted Aloha MAC protocol with  parameter $p$, i.e., every node can access the medium with a probability $p$.

\subsection{NOMA Model}
Several works in NOMA order the receiving nodes by their channel states \cite{ding2014performance}.
However, we consider that the receiving nodes are ordered according to their quality of service (QoS) priorities, since it has been show that it is more realistic assumption \cite{ding2016relay,ding2016mimo}. We consider a scenario in which $D_1$ needs a low data rate but has to be served immediately, whereas $D_2$ needs a higher data rate but can be served later. This can be the case when $D_1$ is a vehicle that needs to receive safety data information about an accident, whereas $D_2$ can be a user that accesses the internet connection.

\subsection{Directional Beamforming Model}

We model the directivity similar to in \cite{singh2015tractable}, where the directional gain, denoted $G(\omega)$, within
the half power beamwidth ($\phi/2$) is $G_{max}$ and is $G_{min}$ in all other
directions. The gain is then expressed as
\begin{equation}
   G(\omega)=\left\{
                \begin{array}{ll}
                  G_{max}, &  \textrm{if} \:\:|\omega|\leq \frac{\phi}{2};\\
                  G_{min}, & \textrm{otherwise}.      
                \end{array}
              \right.
\end{equation}
In this paper, we consider a perfect beam alignment between the nodes, hence $G_{eq} = G_{max}^2$. The impact of beam misalignment is beyond the scope of this paper, and can be a topic of future
works. 

\subsection{Channel and Interference Model}

 We consider an interference limited scenario, that is, the power of noise is set to zero ($\sigma^2=0$). Without loss of generality, we assume  that all nodes transmit with a unit power. 
The signal transmitted by $S$, denoted $ \chi_{S}$ is a mixture of the message intended to $D_1$ and $D_2$. This can be expressed as
\begin{equation}
 \chi_{S}=\sqrt{a_1}\chi_{D_1}+\sqrt{a_2}\chi_{D_2}, \nonumber
 \end{equation}
where $a_i$ is the power coefficients allocated to $D_i$, and $\chi_{Di}$ is the message intended to $D_i$, where $i \in \{1,2\}$. Since $D_1$ has higher power than $D_2$, that is $a_1 \ge a_2$, then $D_1$ comes first in the decoding order. Note that, $a_1+a_2=1$.\\

 The signal received at $D_i$ is expressed as
 \begin{multline}
   \mathcal{Y}_{D_i}=h_{SD_i}\sqrt{r_{SD_i}^{-\alpha}\Upsilon}\:\chi_{S}\mathds{1}(\textrm{LOS}_{SD_i})
   + h_{SD_i}\sqrt{r_{SD_i}^{-\alpha}\Upsilon}\:\chi_{S}\mathds{1}(\textrm{NLOS}_{SD_i})
+ \sum_{x\in \Phi^{\textrm{LOS}}_{X_{D_i}}}h_{D_ix}\sqrt{r_{D_{i}x}^{-\alpha_{\textrm{LOS}}}\Upsilon}\:\chi_{x} \\
 +\sum_{y\in \Phi^{\textrm{LOS}}_{Y_{D_i}}}h_{D_iy}\sqrt{r_{D_{i}y}^{-\alpha_{\textrm{LOS}}}\Upsilon}\:\chi_{y}
 +\sum_{x\in \Phi^{\textrm{NLOS}}_{X_{D_i}}}h_{D_ix}\sqrt{r_{D_{i}x}^{-\alpha_{\textrm{NLOS}}}\Upsilon}\:\chi_{x} 
 +\sum_{y\in \Phi^{\textrm{NLOS}}_{Y_{D_i}}}h_{D_iy}\sqrt{r_{D_{i}y}^{-\alpha_{\textrm{NLOS}}}\Upsilon}\:\chi_{y}, \nonumber
 \end{multline}
where $\mathcal{Y}_{D_i}$ is the signal received by $D_i$, and $\chi_{S}$ is the message transmitted by $S$.
The messages transmitted by the interfere node $x$ and $y$, are denoted respectively by $ \chi_x$ and $\chi_y $. The term $\Upsilon= G_{eq} \eta^2/ (4 \pi)^2$ models the directional gain, the reference path loss at one meter, and $\eta$ is the wavelength of the operating frequency.

The coefficient $h_{SD_i}$ denotes the fading of the link $S-D_i$. The fading coefficient $h_{SD_i}$ is distributed according to a Nakagami-$m$ distribution with parameter $m$ \cite{belbase2018coverage}, that is 
\begin{equation}
f_{h_{SD_i}}(x)=2 \Big(\frac{m}{\mu}\Big)^m \frac{x^{2m-1}}{\Gamma(m)}e^{-\frac{m}{\mu}x^2},
\end{equation}
where the parameter $m \in \{m_{\textrm{LOS}}, m_{\textrm{NLOS}}\}$. Note that $m=m_{\textrm{LOS}}$ when $S-D_i$ is in a LOS, whereas $m=m_{\textrm{NLOS}}$, when $S-D_i$ is in a NLOS. The parameter $\mu$ is the average
received power.
Hence, the power fading coefficient $|h_{SD_i}|^2$ is distributed according to a gamma distribution, that is, 
\begin{equation}
f_{| h_{SD_i} | ^2}(x)=\Big(\frac{m}{\mu}\Big)^m \frac{x^{m-1}}{\Gamma(m)}e^{-\frac{m}{\mu}x}.
\end{equation}

The fading coefficients $h_{D_{i}x}$ and  $h_{D_iy}$ denote the fading of the link $D_i-x$, and $D_i-y$. The fading coefficients are modeled as Rayleigh fading \cite{deng2017meta}. Thus, the power fading coefficients $|h_{D_ix}|^2$ and $|h_{D_iy}|^2$, are distributed according to an exponential distribution with unit mean.

The aggregate interference from the $X$ road at ${D_i}$, denoted $I_{X_{D_i}}$, is expressed as
\begin{eqnarray}\label{EQ.1}
I_{X_{D_i}}&=& I^{\textrm{LOS}}_{X_{D_i}} +I^{\textrm{NLOS}}_{X_{D_i}}  \nonumber \\
&=&\sum_{x\in \Phi^{\textrm{LOS}}_{X_{D_i}}}\vert h_{{D_i}x}\vert^{2}r_{{D_i}x}^{-\alpha_{\textrm{LOS}}}\Upsilon + \sum_{y\in \Phi^{\textrm{NLOS}}_{X_{{D_i}}}}\vert h_{{D_i}x}\vert^{2}r_{{D_i}x}^{-\alpha_{\textrm{NLOS}}}\Upsilon, 
\end{eqnarray}
where $I^{\textrm{LOS}}_{X_{D_i}} $ denotes the aggregate interference from the $X$ road that are in a LOS with ${D_i}$, and $I^{\textrm{NLOS}}_{X_{{D_i}}} $ denotes the aggregate interference from the $X$ road that are in a NLOS with ${D_i}$. 
Similarly, $\Phi^{\textrm{LOS}}_{X_{D_i}}$ and $\Phi^{\textrm{NLOS}}_{X_{D_i}}$, denote respectively, the set of the interferers from the $X$ road at ${D_i}$ in a LOS, and in NLOS.
In the same way, the aggregate interference from the $Y$ road at ${D_i}$, denoted $I_{Y_{D_i}}$, is expressed as
\begin{eqnarray}\label{EQ.2}
I_{Y_{D_i}}&=&  I^{\textrm{LOS}}_{Y_{{D_i}}} + I^{\textrm{NLOS}}_{Y_{{D_i}}}\nonumber \\ 
&=& \sum_{y\in \Phi^{\textrm{LOS}}_{Y_{{D_i}}}}\vert h_{{D_i}y}\vert^{2}r_{{D_i}y}^{-\alpha_{\textrm{LOS}}}\Upsilon +\sum_{y\in \Phi^{\textrm{NLOS}}_{Y_{{D_i}}}}\vert h_{{D_i}y}\vert^{2}r_{{D_i}y}^{-\alpha_{\textrm{NLOS}}}\Upsilon, 
\end{eqnarray}
where $I^{\textrm{LOS}}_{Y_{{D_i}}} $ denotes the aggregate interference from the $X$ road that are in a LOS with ${D_i}$, and $I^{\textrm{NLOS}}_{Y_{{D_i}}} $ denotes the aggregate interference from the $Y$ road that are in a NLOS with ${D_i}$. 
Similarly, $\Phi^{\textrm{LOS}}_{Y_{{D_i}}}$ and $\Phi^{\textrm{NLOS}}_{Y_{{D_i}}}$, denote respectively, the set of the interferers from the $Y$ road at ${D_i}$ in a LOS, and in NLOS.

\section{NOMA Outage Expressions}
\subsection{Signal-to-Interference Ratio (SIR) Expressions}
The outage probability is defined as the probability that the SIR at the receiver is below a given threshold. According to successive interference cancellation (SIC) \cite{hasna2003performance}, $D_1$ message is decoded first at the receiver since it has the higher power allocation, and $D_2$ message is considered as interference. 
The SIR at $D_1$ to decode its desired message, denoted $\textrm{SIR}^{(\alpha)}_{D_1}$, is given by

\begin{equation}\label{EQ.5}
\textrm{SIR}^{(\alpha)}_{D_1}=\frac{\vert h_{SD_1}\vert^{2}r_{SD_1}^{-\alpha}\Upsilon \,a_1}{\vert h_{SD_1}\vert^{2}r_{SD_1}^{-\alpha}\Upsilon a_2+I_{X_{D_1}}+I_{Y_{D_1}}} .
\end{equation}
In order for $D_2$ to decode its desired message, it has to decode $D_1$ message. The SIR at $D_2$ to decode $D_1$ message, denoted $\textrm{SIR}^{(\alpha)}_{D_{2-1}}$, is expressed as
\begin{equation}\label{EQ.6}
\textrm{SIR}^{(\alpha)}_{D_{2-1}}=\frac{\vert h_{SD_2}\vert^{2}r_{SD_2}^{-\alpha}\Upsilon \,a_1}{\vert h_{SD_2}\vert^{2}r_{SD_2}^{-\alpha} \Upsilon a_2+I_{X_{D_2}}+I_{Y_{D_2}}}.
\end{equation}
The SIR at $D_2$ to decode its desired message, denoted $\textrm{SIR}^{(\alpha)}_{D_{2-2}}$, is expressed as 
\begin{equation}\label{EQ.7}
\textrm{SIR}^{(\alpha)}_{D_{2-2}}=\frac{\vert h_{SD_2}\vert^{2}r_{SD_2}^{-\alpha}\Upsilon \,a_2}{I_{X_{D_2}}+I_{Y_{D_2}}}.
\end{equation}
\subsection{Outage Event Expressions}

The outage event that $D_1$ does not decode its desired message, denoted $\textit{O}_{D_1}$, is defined as
\begin{equation}\label{EQ.9}
\mathcal{O}_{D_1}\triangleq \bigcup_{{\textrm{Z}}\in \{\textrm{LOS},\textrm{NLOS}\}} \big\{ \textrm{Z}_{SD_1} \cap (\textrm{SIR}^{(\alpha_{\textrm{Z}})}_{D_1}< \Theta_1) \big\} ,  
\end{equation}
where $\Theta_1=2^{2\mathcal{R}_1}-1$, and $\mathcal{R}_1$ is the target data rate of $D_1$.

Also, the outage event that $D_2$ does not decode its desired message, denoted $\textit{O}_{D_2}$, is defined as
\begin{equation}\label{EQ.12}
\mathcal{O}_{D_2}\triangleq \bigcup_{{\textrm{Z}}\in \{\textrm{LOS},\textrm{NLOS}\}}\bigcup_{i=1}^{2} \big\{ \textrm{Z}_{SD_2} \cap (\textrm{SIR}^{(\alpha_{\textrm{Z}})}_{D_{2-i}}< \Theta_i) \big\} ,
\end{equation}
where $\Theta_2=2^{2\mathcal{R}_2}-1$ ($i=2)$, and $\mathcal{R}_2$ is the target data rate of $D_2$.

\subsection{Outage Probability Expressions}
The outage probability related to $\mathcal{O}_{D_1}$, denoted $\mathbb{P}(\mathcal{O}_{D_1})$, is given when $\Theta_1 < \frac{a_1}{a_2}$, by  
\begin{multline}\label{EQ.14}
\mathbb{P}(\mathcal{O}_{D_1})=1-
 \sum_{\textrm{Z}\in\{\textrm{LOS},\textrm{NLOS}\}}^{} \mathbb{P}(\textrm{Z}_{SD_1})  \\
 \prod_{\textrm{K}\in\{\textrm{LOS},\textrm{NLOS}\}} \sum_{k=0}^{m_\textrm{Z}-1}\frac{1}{k!}\:\big(-\dfrac{m_\textrm{Z}\:\Psi_1}{\mu \:r_{SD_1}^{-\alpha_\textrm{Z}} \Upsilon}\big)^k \sum_{n=0}^{k}\binom{k}{n}\frac{\textrm{d}^{k-n} \mathcal{L}_{I^{\textrm{K}}_{X_{D_1}}}\big(\dfrac{m_\textrm{Z}\:\Psi_1}{\mu \:r_{SD_1}^{-\alpha_\textrm{Z}}\Upsilon }\big)}{\textrm{d}^{k-n} \big(\dfrac{m_\textrm{Z}\:\Psi_1}{\mu \:r_{SD_1}^{-\alpha_\textrm{Z}}\Upsilon }\big)} \frac{\textrm{d}^{n} \mathcal{L}_{I^{\textrm{K}}_{Y_{D_1}}}\big(\dfrac{m_\textrm{Z}\:\Psi_1}{\mu \:r_{SD_1}^{-\alpha_\textrm{Z}}\Upsilon }\big)}{\textrm{d}^{n} \big(\dfrac{m_\textrm{Z}\:\Psi_1}{\mu \:r_{SD_1}^{-\alpha_\textrm{Z}} \Upsilon}\big)},
\end{multline}
where $\Psi_{1}=\Theta_1/(a_1-\Theta_1a_2)$.
\begin{figure}[]
\centering
\includegraphics[scale=0.45]{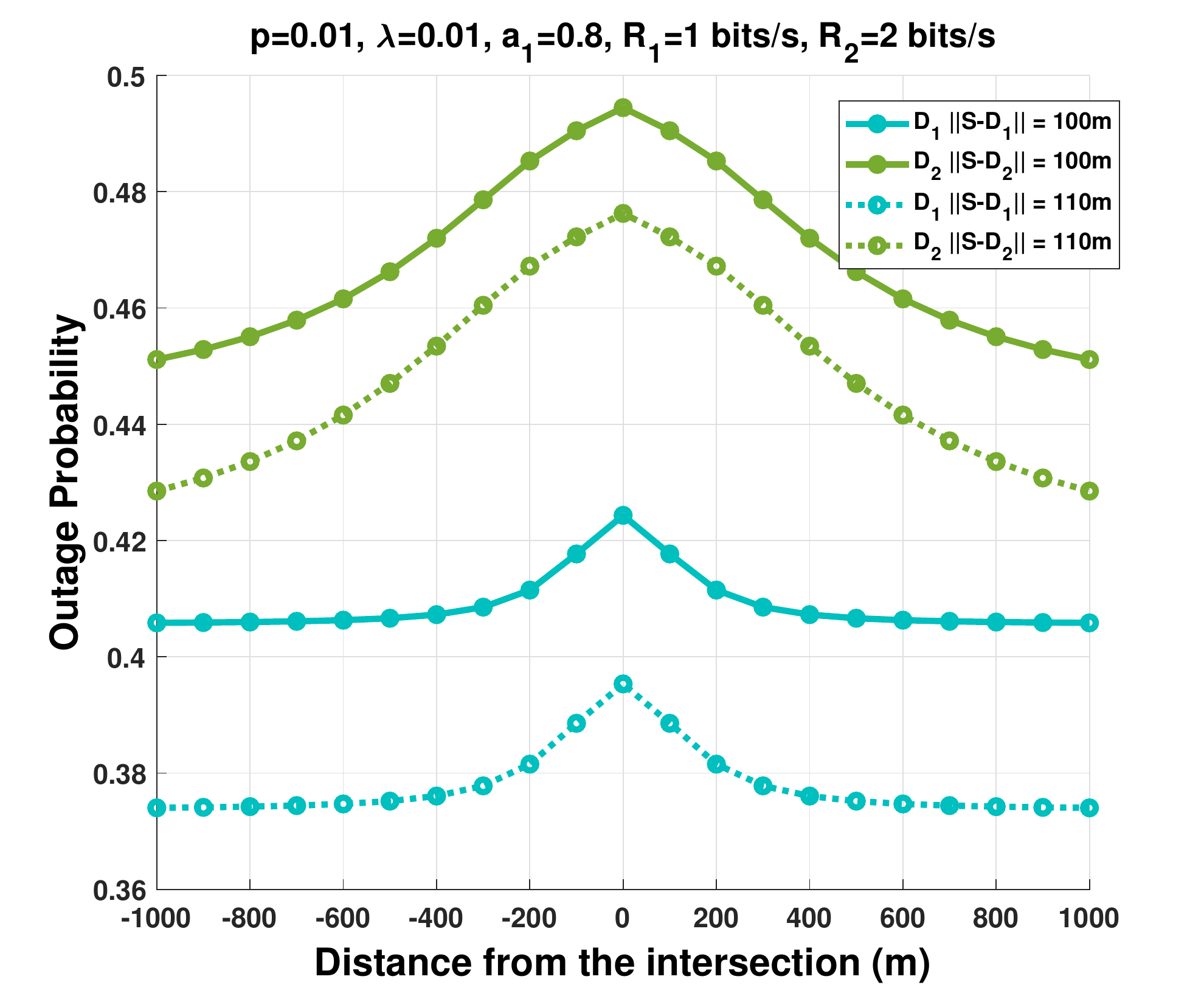}
\caption{Outage probability as a function of distance from the intersection.}
\label{Fig2}
\end{figure}
\begin{figure}[]
\centering
\includegraphics[scale=0.52]{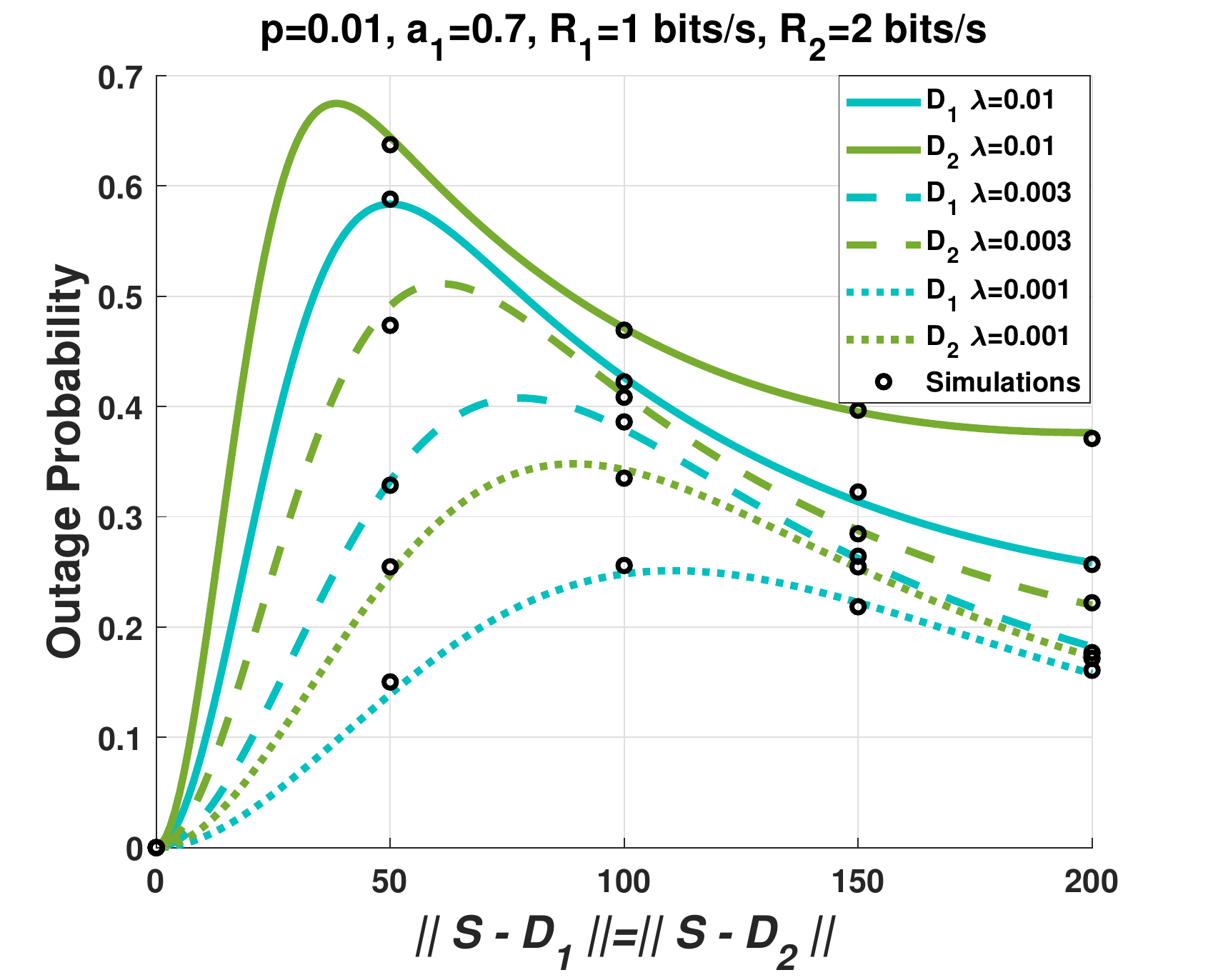}
\caption{Outage probability as a function of $\Vert S -D_1\Vert  =  \Vert S -D_2\Vert  $. }
\label{Fig3}
\end{figure}
\begin{figure}[]
\centering
\includegraphics[scale=0.45]{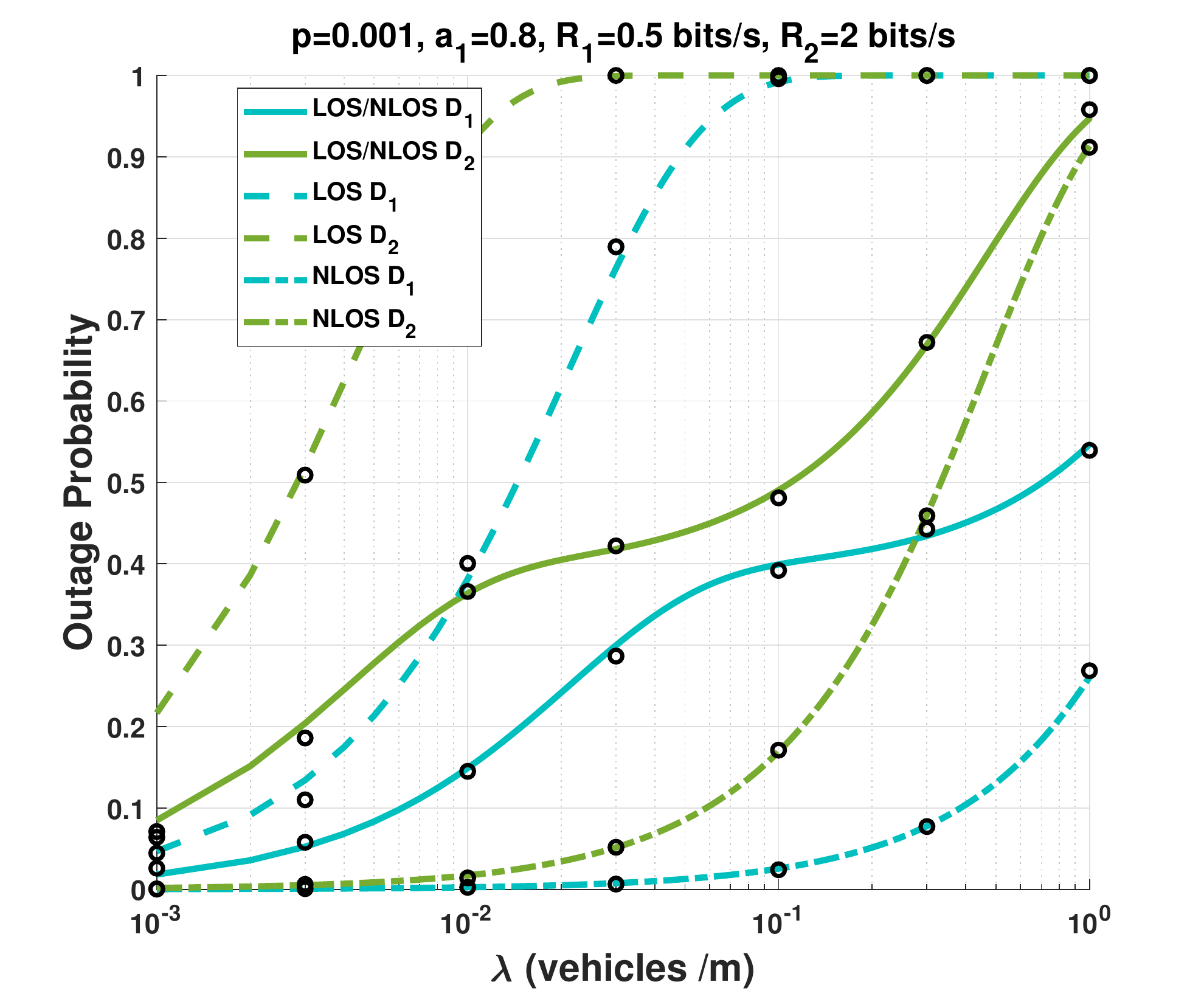}
\caption{Outage probability as function of $\lambda$ for LOS transmission, NLOS, and LOS/NLOS (equation (\ref{EQ.14}) and (\ref{EQ.15})).}
\label{Fig4}
\end{figure}

The outage probability related to $\mathcal{O}_{D_2}$, denoted by $\mathbb{P}(\mathcal{O}_{D_2})$ is given, when $\Theta_1 < \frac{a_1}{a_2}$, by  
\begin{multline}\label{EQ.15}
\mathbb{P}(\mathcal{O}_{D_2})=1-
 \sum_{\textrm{Z}\in\{\textrm{LOS},\textrm{NLOS}\}}^{} \mathbb{P}(\textrm{Z}_{SD_2}) \\ 
 \prod_{\textrm{K}\in\{\textrm{LOS},\textrm{NLOS}\}} \sum_{k=0}^{m_\textrm{Z}-1}\frac{1}{k!}\:\big(-\dfrac{m_\textrm{Z}\:\Psi_{\mathrm{max}}}{\mu \:r_{SD_2}^{-\alpha_\textrm{Z}} \Upsilon}\big)^k \sum_{n=0}^{k}\binom{k}{n}\frac{\textrm{d}^{k-n} \mathcal{L}_{I^{\textrm{K}}_{X_{D_2}}}\big(\dfrac{m_\textrm{Z}\:\Psi_{\mathrm{max}}}{\mu \:r_{SD_2}^{-\alpha_\textrm{Z}}\Upsilon }\big)}{\textrm{d}^{k-n} \big(\dfrac{m_\textrm{Z}\:\Psi_{\mathrm{max}}}{\mu \:r_{SD_2}^{-\alpha_\textrm{Z}}\Upsilon }\big)} \frac{\textrm{d}^{n} \mathcal{L}_{I^{\textrm{K}}_{Y_{D_2}}}\big(\dfrac{m_\textrm{Z}\:\Psi_{\mathrm{max}}}{\mu \:r_{SD_2}^{-\alpha_\textrm{Z}}\Upsilon }\big)}{\textrm{d}^{n} \big(\dfrac{m_\textrm{Z}\:\Psi_{\mathrm{max}}}{\mu \:r_{SD_2}^{-\alpha_\textrm{Z}} \Upsilon}\big)},
\end{multline}
where $\Psi_{\mathrm{max}}=\mathrm{max}(\Psi_1,\Psi_2)$, and $\Psi_2=\Theta_2/a_2$.\\
\textit{Proof}:  See Appendix \ref{App_A}.\hfill $ \blacksquare $ \\

 \section{Laplace Transform Expressions}
 
In this section, we present the Laplace transform expressions of the interference from the $X$ road at $D_i$, denoted $\mathcal{L}_{I^{\textrm{K}}_{X_{D_i}}}$, and from the $Y$ road at $D_i$, denoted $\mathcal{L}_{I^{\textrm{K}}_{Y_{D_i}}}$. 

The Laplace transform of the interference originating from the $X$ road at the received node denoted  $D_i$, is expressed as

\begin{equation}\label{eq:33}
\mathcal{L}_{I^{\textrm{K}}_{X_{D_i}}}(s)=\exp\Bigg(-\emph{p}\lambda^{\textrm{K}}_{X}\int_\mathbb{R}\dfrac{1}{1+\Vert \textit{x}-{D_i} \Vert^{\alpha_{\textrm{K}}}/s}\textrm{d}x\Bigg),
\end{equation}
where 
\begin{equation}\label{eq:34}
\Vert \textit{x}-{D_i} \Vert=\sqrt{\big[d_i\sin(\theta_{{D_i}})\big]^2+\big[x-d_i \cos(\theta_{D_i}) \big]^2 }.
\end{equation}

The Laplace transform of the interference originating from the $Y$ road at $D_i$ is given by
 
\begin{equation}\label{eq:35}
\mathcal{L}_{I^{\textrm{K}}_{Y_{D_i}}}(s)=\exp\Bigg(-\emph{p}\lambda^{\textrm{K}}_{Y}\int_\mathbb{R}\dfrac{1}{1+\Vert \textit{y}-{D_i} \Vert^{\alpha_{\textrm{K}}}/s}\textrm{d}y\Bigg),
\end{equation}
where
\begin{equation}\label{eq:36}
\Vert \textit{y}-{D_i} \Vert=\sqrt{\big[d_i\cos(\theta_{{D_i}})\big]^2+\big[y-d_i \sin(\theta_{D_i}) \big]^2 },
\end{equation}
where $\theta_{D_i} $ is the angle between the node ${D_i}$ and the $X$ road.
\textit{Proof}:  See Appendix \ref{App_B}.\hfill $ \blacksquare $ \\

We only present the case when $\alpha_{\textrm{K}}=2$ due to the lack of space. The Laplace transform expressions of the interference at $D_i$ for an intersection scenario, when  $\alpha_{\textrm{K}}=2$ are given by 
\begin{equation}\label{EQ.17}
\mathcal{L}_{I^{\textrm{K}}_{X_{D_i}}}(s)=\exp\Bigg(\dfrac{-\emph{p}\lambda^{\textrm{K}}_{X}s\pi}{\sqrt{\big[{d_i}\sin(\theta_{{D_i}})\big]^2+s}}\Bigg),
\end{equation}
and
\begin{equation}\label{EQ.18}
\mathcal{L}_{I^{\textrm{K}}_{Y_{D_i}}}(s)=\exp\Bigg(\dfrac{-\emph{p}\lambda^{\textrm{K}}_{Y}s\pi}{\sqrt{\big[{d_i}\cos(\theta_{{D_i}})\big]^2+s}}\Bigg).
\end{equation}
\textit{Proof}:  See Appendix \ref{App_C}.\hfill $ \blacksquare $ \\

\begin{figure}[]
\centering
\includegraphics[scale=0.45]{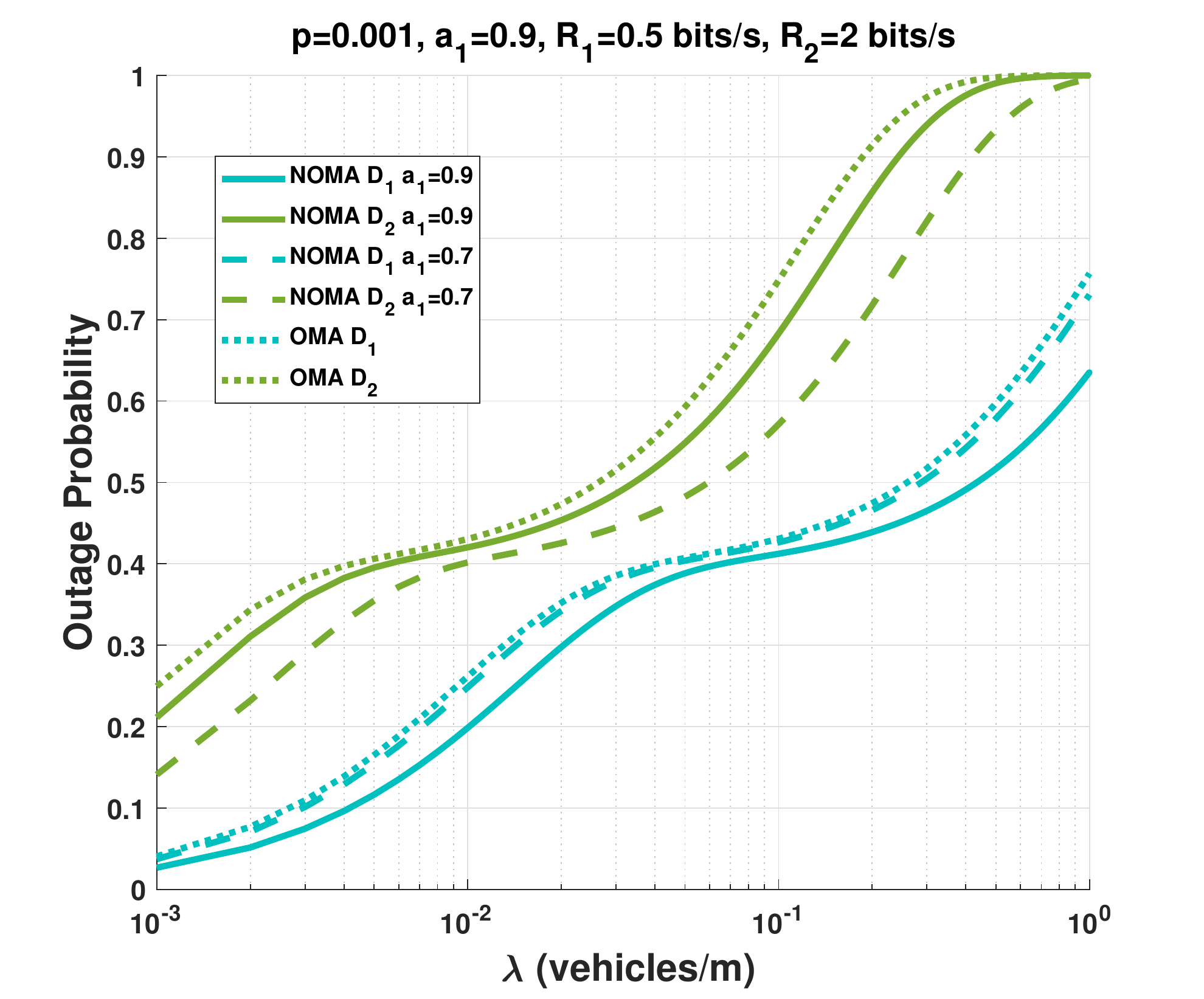}
\caption{Outage probability as a function of $\lambda$ considering NOMA and OMA.}
\label{Fig5}
\end{figure}
\begin{figure}[]
\centering
\includegraphics[scale=0.45]{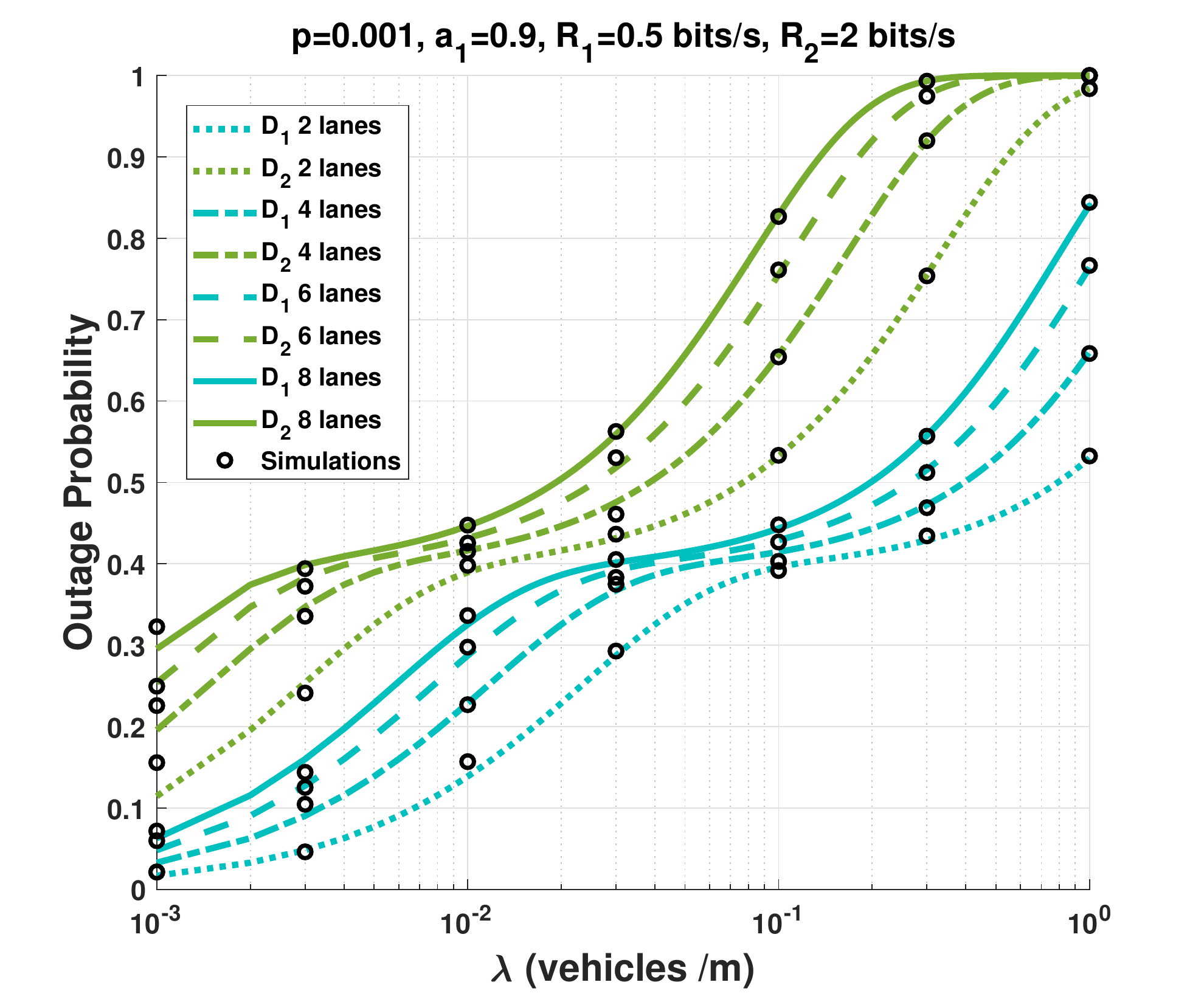}
\caption{Outage probability as a function of $\lambda$ for several values of the number of lanes.}
\label{Fig6}
\end{figure}
\section{Simulations and Discussions}

In this section, we evaluate the performance of mmWave VCs using NOMA at road intersections. 
In order to verify the accuracy of the theoretical results, Monte-Carlo simulations are carried out by averaging over 10,000 realizations of the PPPs and fading parameters. Monte Carlo simulations are carried out, and they match perfectly the theoretical results, which validates the correctness of our analysis. We set, without loss of generality,  $\lambda^{\textrm{LOS}}_X = \lambda^{\textrm{LOS}}_Y = \lambda^{\textrm{NLOS}}_X = \lambda^{\textrm{NLOS}}_Y =\lambda$. $S=(0,0)$, $D_1=(100,10)$, and $D_2=(100,-10)$, $\beta=9.5\times10^3$ \cite{bai2014analysis}, $\mu=1$. We set $\alpha_{\textrm{LOS}}=2$, $\alpha_{\textrm{NLOS}}=4$, $m_{\textrm{LOS}}=2$,  and  $m_{\textrm{NLOS}}=1$. Finally, we set $G_{max}=18$ dBi, $\eta=30$ GHz. Unless stated otherwise, we consider mmWave VCs using NOMA in all the results.

Fig.\ref{Fig2} shows the outage probability as a function of the distance of the triplet $\{S,D_1,D_2\}$ from the intersection.
We see from Fig.\ref{Fig2} that the outage probability increases as the vehicles drive toward the intersection. This is because when the vehicles are far from the intersection, only the interferes in the same road segment contribute to the aggregate interference. But, as the vehicles approach the intersection, both road segments contribute to the aggregate interference. We can also see that the outage probability when $\Vert S-D_1 \Vert=100$ m ($\Vert S-D_2 \Vert=100$ m) is higher than  the outage probability when $\Vert S-D_1 \Vert=110$ m ($\Vert S-D_2 \Vert=110$ m). This result is counter-intuitive because large distance decreases the path loss, and thus, it decreases the SIR at the receiver and increases the outage probability.

To investigate the effect of the distance between the transmitter and the receiver, Fig.\ref{Fig3} depicts the outage probability as a function of the distance between the source and the destinations.
We can see from Fig.\ref{Fig3} that the outage probability increases, as the distance between the source and the destinations increases, until it reaches its maximum point and then it deceases.
This because, as the distance between the transmitting and the receiving nodes increases, the LOS probability decreases, and the NLOS probability increases, which decreases SIR at the receiver, and hence, decreases the outage probability. 
We can also see that the peak of the outage probability is reached for different values of $\lambda$.
In fact, we can see that for high values of $\lambda$, the peak is reached for short distances ($\Vert S-D_1\Vert = 50$ m and  $\Vert S-D_2\Vert=40$ m). However, for low values of $\lambda$, the peak is reached for high distances ($\Vert S-D_1\Vert = 110$ m and  $\Vert S-D_2\Vert=90$ m).

To show the effect of LOS and NLOS on the performance, Fig.\ref{Fig4} plots the outage probability as function of $\lambda$ for LOS transmission, NLOS transmission, and LOS/NLOS, that is equation (\ref{EQ.14}) and (\ref{EQ.15}).
We can see that LOS scenario has the highest outage probability. This is because, when the interference are in direct line of sight with $D_1$ and $D_2$, the power of aggregate interference increases, hence reducing the SIR and increasing the outage probability. On the other hand, the NLOS scenario has the smallest outage probability, since the interference are in NLOS with the transmitting nodes. The model for this paper include a blockage model that includes both LOS and NLOS. Therefore, we wan see that the performance are between the LOS scenario and NLOS scenario, which are two extreme cases.

 Fig.\ref{Fig5} compares the outage probability of mmWave VCs using NOMA and OMA. We can see that NOMA outperforms OMA. We can also see that when $a_1=0,9$, the outage probability of $D_1$ decreases, and the outage probability of $D_2$ increases. However, when $a_1=0,7$, the outage probability of $D_1$ increases, and the outage probability of $D_2$ decreases. This because for low values of $a_1$, there is more power allocated to $D_2$, which increases its SIR and decreases its outage probability. On the other hand, high values of $a_1$, means more power is allocated to $D_1$, which increases its SIR and its decreases its outage probability.

Now, we investigate a more realistic intersection scenario involving several lanes.
Fig.\ref{Fig6} plots the outage probability as a function of $\lambda$ for several values of the number of lanes. We can see from Fig.\ref{Fig6} that the outage probability increases as the number of lanes increases. This because, when the number of lanes increases, the number of interfering vehicles increases as well, which increases the aggregate interference at the receiver, hence, increasing the outage probability.

\section{Conclusion}
 In this paper, we studied performance of mmWave VCs using NOMA at road intersections. We studied the case when the intersection involves of two perpendicular lanes, and then extended the study to an intersection with several lanes. 
 Our analysis included the effects of blockage from buildings and vehicles at intersections. 
 Closed form outage probability expressions were obtained.
We showed that as the nodes reach the intersection, the outage probability increases. This makes intersections a critical areas.
We also showed that that the peak of the outage probability is reached for different values of $\lambda$.
We showed that for high density scenarios (high values of $\lambda$), the peak is reached for short distances, whereas for low density scenarios, the peak is reached for high distances.
Counter-intuitively, we showed that NLOS scenario has a better performance than LOS scenario. 
 Finally, we compared the performance of mmWave NOMA with OMA, and we showed that OMA offers a significant improvement over cooperative OMA.

\appendices
\section{}\label{App_A}

To calculate $\mathbb{P}(\mathcal{O}_{D_1})$, we express as follows
\begin{equation} \label{App.1}
\mathbb{P}(\mathcal{O}_{D_1})=1-\mathbb{P}(\mathcal{O}^{C}_{D_1}).
\end{equation}
The probability $\mathbb{P}(\mathcal{O}^{C}_{D_1})$ is expressed as
\begin{eqnarray}\label{App.8}
\mathbb{P}(\mathcal{O}^{C}_{D_1})&=&
\sum^{}_{{\textrm{Z}}\in \{\textrm{LOS},\textrm{NLOS}\}} \mathbb{E}_{I_{X},I_Y}\Bigg[\mathbb{P}\Bigg\lbrace\ \textrm{Z}_{SD_1
}\cap (\textrm{SIR}^{(\alpha_{\textrm{Z}})}_{D_1} \geq \Theta_1)\Bigg\rbrace\Bigg]\nonumber  \\
&=&\sum^{}_{{\textrm{Z}}\in \{\textrm{LOS},\textrm{NLOS}\}}\mathbb{P}(\textrm{Z}_{SD_1})\:\:     \mathbb{E}_{I_{X},I_Y}\Bigg[\mathbb{P}\Bigg\lbrace\  \textrm{SIR}^{(\alpha_{\textrm{Z}})}_{D_1} \geq \Theta_1\Bigg\rbrace\Bigg] \nonumber \\
&=&\sum^{}_{{\textrm{Z}}\in \{\textrm{LOS},\textrm{NLOS}\}}\mathbb{P}(\textrm{Z}_{SD_1})\:\:     \mathbb{E}_{I_{X},I_Y}\Bigg[\mathbb{P}\Bigg\lbrace\ \frac{\vert h_{SD_1}\vert^{2}r_{SD_1}^{-\alpha_{\textrm{Z}}}\Upsilon a_1}{\vert h_{SD_1}\vert^{2}r_{SD_1}^{-\alpha_{{\textrm{Z}}}}\Upsilon a_2+I_{X_{D_1}}+I_{Y_{D_1}}} \ge \Theta_1\Bigg\rbrace\Bigg]\nonumber  \\
&=&\sum^{}_{{\textrm{Z}}\in \{\textrm{LOS},\textrm{NLOS}\}}\mathbb{P}(\textrm{Z}_{SD_1})\:\:   \mathbb{E}_{I_{X},I_Y}\Bigg[\mathbb{P}\Bigg\lbrace\vert h_{SD_1}\vert^{2}r_{SD_1}^{-\alpha_{\textrm{Z}}} \Upsilon (a_1-\Theta_1 a_2)\ge
\Theta_1\big[I_{X_{D_1}}+I_{Y_{D_1}}\big]\Bigg\rbrace\Bigg].\nonumber \\ 
\end{eqnarray}
We can notice from (\ref{App.8}) that, when $\Theta_1 \ge a_1/ a_2$, $\mathbb{P}(\mathcal{O}^{C}_{D_1})=0$. Then, when $\Theta_1 < a_1/ a_2$, and after setting $\Psi_{1}= \Theta_1 /(a_1- \Theta_1 a_2)$, we get
\begin{equation}\label{App.9}
\mathbb{P}(\mathcal{O}^{C}_{D_1})=\sum^{}_{{\textrm{Z}}\in \{\textrm{LOS},\textrm{NLOS}\}}\mathbb{P}(\textrm{Z}_{SD_1})\:\:   \mathbb{E}_{I_{X},I_Y}\Bigg[\mathbb{P}\Bigg\lbrace\vert h_{SD_1}\vert^{2}\ge\frac{\Psi_{1}}{r_{SD_1}^{-\alpha_{\textrm{Z}} }\Upsilon  }\big[I_{X_{D_1}}+I_{Y_{D_1}}\big]\Bigg\rbrace\Bigg].
\end{equation}
Since $\vert h_{SD_1}\vert^{2}$ follows a gamma distribution, its complementary cumulative distribution function (CCDF) is given by
\begin{equation}\label{App.10}
\bar{F}_{\vert h_{SD_1}\vert^{2}}(X)=\mathbb{P}(\vert h_{SD_1}\vert^{2}>X)=\frac{\Gamma(m_{\textrm{Z}},\frac{m_{\textrm{Z}}}{\mu }X)}{\Gamma(m_{\textrm{Z}})},
\end{equation}
hence (\ref{App.9}) becomes 
\begin{eqnarray}
\mathbb{P}(\mathcal{O}^{C}_{D_1})&=&\sum^{}_{{\textrm{Z}}\in \{\textrm{LOS},\textrm{NLOS}\}}\mathbb{P}(\textrm{Z}_{SD_1})\nonumber\\
&&\mathbb{E}_{I_{X},I_Y}\Bigg[\frac{\Gamma\Big(m_{\textrm{Z}},\dfrac{m_{\textrm{Z}} \:\Psi_1}{\mu\: r_{SD_1}^{-\alpha_{{\textrm{Z}}}} \Upsilon } (I^{\textrm{LOS}}_{X_{D_1}}+I^{\textrm{LOS}}_{Y_{D_1}})\Big)}{\Gamma(m_{\textrm{Z}})}\Bigg] \times 
\mathbb{E}_{I_{X},I_Y}\Bigg[\frac{\Gamma\Big(m_{\textrm{Z}},\dfrac{m_{\textrm{Z}} \:\Psi_1}{\mu\: r_{SD_1}^{-\alpha_{{\textrm{Z}}}} \Upsilon } (I^{\textrm{NLOS}}_{X_{D_1}}+I^{\textrm{NLOS}}_{Y_{D_1}})\Big)}{\Gamma(m_{\textrm{Z}})}\Bigg]\nonumber.
\end{eqnarray}
Therefore, we obtain
\begin{equation}\label{App.11}
\mathbb{P}(\mathcal{O}^{C}_{D_1})=\sum^{}_{{\textrm{Z}}\in \{\textrm{LOS},\textrm{NLOS}\}}\mathbb{P}(\textrm{Z}_{SD_1})\:\:   \prod^{}_{{\textrm{K}}\in \{\textrm{LOS},\textrm{NLOS}\}} \mathbb{E}_{I_{X},I_Y}\Bigg[\frac{\Gamma\Big(m_{\textrm{Z}},\dfrac{m_{\textrm{Z}} \:\Psi_1}{\mu\: r_{SD_1}^{-\alpha_{{\textrm{Z}}}} \Upsilon } (I^{\textrm{K}}_{X_{D_1}}+I^{\textrm{K}}_{Y_{D_1}})\Big)}{\Gamma(m_{\textrm{Z}})}\Bigg].
\end{equation}
The exponential sum function when $m_{\textrm{Z}}$ is an integer is defined as 
\begin{equation}\label{App.12}
e_{(m_{\textrm{Z}})}=\sum_{k=0}^{m_{\textrm{Z}}-1} \frac{(\frac{m_{\textrm{Z}}}{\mu }X)^k}{k!}=e^X \frac{\Gamma(m_{\textrm{Z}},\frac{m_{\textrm{Z}}}{\mu }X)}{\Gamma(m_{\textrm{Z}})},
\end{equation}
then
\begin{equation}\label{App.13}
\frac{\Gamma(m_{\textrm{Z}},\frac{m_{\textrm{Z}}}{\mu }X)}{\Gamma(m_{\textrm{Z}})}=e^{-\frac{m_{\textrm{Z}}}{\mu}X}\sum_{k=0}^{m_{\textrm{Z}}-1}\frac{1}{k!}{\big(\frac{m_{\textrm{Z}}\:X}{\mu}\big)}^k.
\end{equation}
We denote the expectation in equation (\ref{App.11}) by $\mathcal{E}(I_{X},I_Y)$, then $\mathcal{E}(I_{X},I_Y)$ equals
\begin{eqnarray}\label{App.14}
\mathcal{E}(I_{X},I_Y)&=&\mathbb{E}_{I_{X},I_Y}\Bigg[\exp\Big(-\dfrac{m_{\textrm{Z}} \:\Psi_1}{\mu\: r_{SD_1}^{-\alpha_{\textrm{Z}}}\Upsilon } (I^{\textrm{K}}_{X_{D_1}}+I^{\textrm{K}}_{Y_{D_1}})\Big)
\ \sum_{k=0}^{m_{\textrm{Z}}-1}\frac{1}{k!}\Big(\dfrac{m_{\textrm{Z}}\:\Psi_1}{\mu \:r_{SD_1}^{-\alpha_{\textrm{Z}}}\Upsilon } (I^{\textrm{K}}_{X_{D_1}}+I^{\textrm{K}}_{Y_{D_1}})\Big)^k\Bigg]\nonumber\\
&=&\sum_{k=0}^{m-1}\frac{1}{k!}\:\Big(\dfrac{m_{\textrm{Z}}\:\Psi_1}{\mu \:r_{SD_1}^{-\alpha_{\textrm{Z}}}\Upsilon }\Big)^k  \mathbb{E}_{I_{X},I_Y}\Bigg[\exp\Big(-\dfrac{m_{\textrm{Z}} \:\Psi_1}{\mu\: r_{SD_1}^{-\alpha_{\textrm{Z}}}\Upsilon } \big(I^{\textrm{K}}_{X_{D_1}}+I^{\textrm{K}}_{Y_{D_1}}\big)\Big)
\big(I^{\textrm{K}}_{X_{D_1}}+I^{\textrm{K}}_{Y_{D_1}}\big)^k\Bigg].\nonumber\\
\end{eqnarray}
Applying the binomial theorem in (\ref{App.14}), we get
\begin{eqnarray}\label{App.15}
\mathcal{E}(I_{X},I_Y)&=&\sum_{k=0}^{m_{\textrm{Z}}-1}\frac{1}{k!}\:\Big(\dfrac{m_{\textrm{Z}}\:\Psi_1}{\mu \:r_{SD_1}^{-\alpha_{\textrm{Z}}}\Upsilon }\Big)^k \nonumber\\&&\times \mathbb{E}_{I_{X},I_Y}\Bigg[\exp\Big(-\dfrac{m_{\textrm{Z}} \:\Psi_1}{\mu\: r_{SD_1}^{-\alpha_{\textrm{Z}}}\Upsilon } \big[I^{\textrm{K}}_{X_{D_1}}+I^{\textrm{K}}_{Y_{D_1}}\big]\Big)
\sum_{n=0}^{k} \binom{k}{n}{(I^{\textrm{K}}_{X_{D_1}})}^{k-n}\:{(I^{\textrm{K}}_{Y_{D_1}})}^n\Bigg]
\nonumber\\
&=&\sum_{k=0}^{m_{\textrm{Z}}-1}\frac{1}{k!}\:\Omega^k \mathbb{E}_{I_{X},I_Y}\Bigg[\exp\Big(-\Omega \big[I^{\textrm{K}}_{X_{D_1}}+I^{\textrm{K}}_{Y_{D_1}}\big]\Big) \times
\sum_{n=0}^{k} \binom{k}{n}{(I^{\textrm{K}}_{X_{D_1}})}^{k-n}\:({I^{\textrm{K}}_{Y_{D_1}}})^n\Bigg],
\end{eqnarray}
where $\Omega=\dfrac{m_{\textrm{Z}}\:\Psi_1}{\mu \:r_{SR}^{-\alpha_{\textrm{Z}}} \Upsilon}$. To calculate the expectation in (\ref{App.15}), denoted $\mathcal{T}(I_{X},I_Y)$, we proceed as follows

\begin{eqnarray}\label{App.17}
\mathcal{T}(I_{X},I_Y)&=&\sum_{n=0}^{k}\binom{k}{n} \nonumber\\ 
&\times& \mathbb{E}_{I_{X},I_Y}\Bigg[e^{-\Omega\:I^{\textrm{K}}_{X_{D_1}}}\:e^{-\Omega\:I^{\textrm{K}}_{Y_{D_1}}}{(I^{\textrm{K}}_{X_{D_1}})}^{k-n}\:{(I^{\textrm{K}}_{Y_{D_1}})}^n\Bigg]\nonumber\\
&=&\sum_{n=0}^{k}\binom{k}{n}\mathbb{E}_{I_{X}}\Bigg[e^{-\Omega\:I^{\textrm{K}}_{X_{D_1}}}{(I^{\textrm{K}}_{X_{D_1}})}^{k-n}\Bigg]\nonumber\\
&\times& \mathbb{E}_{I^{\textrm{K}}_{Y_{D_1}}}\Bigg[e^{-\Omega\:I^{\textrm{K}}_{Y_{D_1}}}\:{(I^{\textrm{K}}_{Y_{D_1}})}^n\Bigg] \nonumber\\
&\overset{(a)}{=}&\sum_{n=0}^{k}\binom{k}{n}(-1)^{k-n}\frac{\textrm{d}^{k-n} \mathcal{L}_{I^{\textrm{K}}_{X_{D_1}}}(\Omega)}{\textrm{d}^{k-n} \Omega} (-1)^{n}\frac{\textrm{d}^{n} \mathcal{L}_{I^{\textrm{K}}_{Y_{D_1}}}(\Omega)}{\textrm{d}^{n} \Omega} \nonumber\\
&=&(-1)^{k}\sum_{n=0}^{k}\binom{k}{n}\frac{\textrm{d}^{k-n} \mathcal{L}_{I^{\textrm{K}}_{X_{D_1}}}(\Omega)}{\textrm{d}^{k-n} \Omega} \frac{\textrm{d}^{n} \mathcal{L}_{I^{\textrm{K}}_{Y_{D_1}}}(\Omega)}{\textrm{d}^{n} \Omega}. \nonumber
\end{eqnarray} 
where (a) stems form the following property 
\begin{eqnarray}\label{App.20}
\mathbb{E}_{I}\Big[e^{-\Omega I}{I}^{N}\Big]&=&(-1)^N \frac{\textrm{d}^{N}\mathbb{E}_{I}\Big[e^{-\Omega\:I}{I}^{N}\Big]}{\textrm{d}^{N} \Omega}\nonumber\\
&=&(-1)^N\frac{\textrm{d}^{N} \mathcal{L}_{I}(\Omega)}{\textrm{d}^{N} \Omega}.\nonumber
\end{eqnarray} 
Finally, the expectation becomes
\begin{multline}\label{App.a21}
\mathcal{T}(I_{X},I_Y)=\sum_{k=0}^{m_{\textrm{Z}}-1}\frac{1}{k!}\:\big(-\dfrac{m_{\textrm{Z}}\:\Psi_1}{\mu \:r_{SD_1}^{-\alpha_{\textrm{Z}}}\Upsilon }\big)^k\sum_{n=0}^{k}\binom{k}{n}\frac{\textrm{d}^{k-n} \mathcal{L}_{I^{\textrm{K}}_{X_{D_1}}}\big(\dfrac{m_{\textrm{Z}}\:\Psi_1}{\mu \:r_{SD_1}^{-\alpha_{\textrm{Z}}}\Upsilon }\big)}{\textrm{d}^{k-n} \big(\dfrac{m_{\textrm{Z}}\:\Psi_1}{\mu \:r_{SD_1}^{-\alpha_{\textrm{Z}}} \Upsilon}\big)} \frac{\textrm{d}^{n} \mathcal{L}_{I^{\textrm{K}}_{Y_{D_1}}}\big(\dfrac{m_{\textrm{Z}}\:\Psi_1}{\mu \:r_{SD_1}^{-\alpha_{\textrm{Z}}}\Upsilon }\big)}{\textrm{d}^{n} \big(\dfrac{m_{\textrm{Z}}\:\Psi_1}{\mu \:r_{SD_1}^{-\alpha_{\textrm{Z}}}\Upsilon }\big)}.
\end{multline}
Then plugging (\ref{App.a21}) in (\ref{App.11}) yields (\ref{EQ.14}).
The expression of $\textrm{d}^{k-n} \mathcal{L}_{I^{\textrm{K}}_{X}}(s)/\textrm{d}^{k-n} (s)$ and $\textrm{d}^{n} \mathcal{L}_{I^{\textrm{K}}_{Y}}(s)/\textrm{d}^{n} (s)$ are given by (\ref{eq.40}) and (\ref{eq.41}).

 In the same way we express $\mathbb{P}(\mathcal{O}^{C}_{D_2})$ as
 \begin{equation}\label{App.23}
\mathbb{P}(\mathcal{O}_{D_2})=1-\mathbb{P}(\mathcal{O}^{C}_{D_2}).
\end{equation}
To calculate $\mathbb{P}(\mathcal{O}^{C}_{D_2})$ we proceed as follows
\begin{eqnarray} \label{App.27}
\mathbb{P}(\mathcal{O}^{C}_{D_2})&=&
\sum^{}_{{\textrm{Z}}\in \{\textrm{LOS},\textrm{NLOS}\}} \mathbb{E}_{I_{X},I_Y}\Bigg[\mathbb{P}\Bigg\lbrace\ \bigcap_{i=1}^{2} \big\{ \textrm{Z}_{SD_2} \cap (\textrm{SIR}^{(\alpha_{\textrm{Z}})}_{D_{2-i}} \geq \Theta_i)\Bigg\rbrace\Bigg] \nonumber \\
&=&\sum^{}_{{\textrm{Z}}\in \{\textrm{LOS},\textrm{NLOS}\}}\mathbb{P}(\textrm{Z}_{SD_2})\:\:      \mathbb{E}_{I_{X},I_Y}\Bigg[\mathbb{P}\Bigg\lbrace\ \bigcap_{i=1}^{2}  \textrm{SIR}^{(\alpha_{\textrm{Z}})}_{D_{2-i}} \geq \Theta_i\Bigg\rbrace\Bigg] \nonumber \\
&=&\sum^{}_{{\textrm{Z}}\in \{\textrm{LOS},\textrm{NLOS}\}}\mathbb{P}(\textrm{Z}_{SD_2})\:\:      \mathbb{E}_{I_{X},I_Y}\Bigg[\mathbb{P}\Bigg\lbrace\  \textrm{SIR}^{(\alpha_{\textrm{Z}})}_{D_{2-1}} \geq \Theta_1 \cap \textrm{SIR}^{(\alpha_{\textrm{Z}})}_{D_{2-2}} \geq \Theta_2\Bigg\rbrace\Bigg].
\end{eqnarray}

Following the same steps as for $\mathbb{P}(\mathcal{O}^{C}_{D_1})$, we get
\begin{multline} 
\mathbb{P}(\mathcal{O}^{C}_{D_2}) = \mathbb{E}_{I_{X},I_Y}\Bigg[\mathbb{P}\Bigg\lbrace\frac{\vert h_{SD_2}\vert^{2}r_{SD_2}^{-\alpha_{\textrm{Z}}} \Upsilon a_1}{\vert h_{SD_2}\vert^{2}r_{SD_2}^{-\alpha_{\textrm{Z}}} \Upsilon a_2+I_{X_{D_2}}+I_{Y_{D_2}}} \ge \Theta_1, \frac{\vert h_{SD_2}\vert^{2}r_{SD_2}^{-\alpha_{\textrm{Z}}} \Upsilon a_2}{I_{X_{D_2}}+I_{Y_{D_2}}} \ge \Theta_2\Bigg\rbrace\Bigg].\nonumber
\end{multline}
When $\Theta_1 > a_1/ a_2$, then $\mathbb{P}(\mathcal{O}_{D_2})=1$, 
otherwise we continue the derivation
We set $\Psi_{2}= \Theta_2/a_2$, then
\begin{eqnarray} 
\mathbb{P}(\mathcal{O}^{C}_{D_2})&=&
\mathbb{E}_{I_{X},I_Y}\Bigg[\mathbb{P}\Bigg\lbrace\vert h_{SD_2}\vert^{2}\ge\frac{\Psi_{1}}{r_{SD_2}^{-\alpha_{\textrm{Z}}} \Upsilon }\big[I_{X_{D_2}}+I_{Y_{D_2}}\big], 
\vert h_{SD_2}\vert^{2}\ge\frac{\Psi_{2}}{r_{SD_2}^{-\alpha_{\textrm{Z}}} \Upsilon }\big[I_{X_{D_2}}+I_{Y_{D_2}}\big]\Bigg\rbrace\Bigg] \nonumber\\
&=&\mathbb{E}_{I_{X},I_Y}\Bigg[\mathbb{P}\Bigg\lbrace\vert h_{SD_2}\vert^{2}\ge\frac{\mathrm{max}(\Psi_{1},\Psi_{2})}{r_{SD_2}^{-\alpha_{\textrm{Z}}} \Upsilon }\big[I_{X_{D_2}}+I_{Y_{D_2}}\big]\Bigg\rbrace\Bigg].\nonumber
\end{eqnarray}
Following the same steps as for $\mathbb{P}(\mathcal{O}^{C}_{D_1})$, we obtain (\ref{EQ.15}).
\section{}\label{App_B} 
The Laplace transform of the interference originating from the $X$ road at ${D_i}$ is expressed as 
\begin{eqnarray}\label{eq:64}
\mathcal{L}_{{I^{\textrm{K}}_{X_{D_i}}}}(s) 
&=&\mathbb{E}\Bigg[{\exp\big(-sI^{\textrm{K}}_{X_{D_i}} \big)}\Bigg]\nonumber\\
&=&\mathbb{E}\Bigg[{\exp\Bigg(-\sum_{x\in\Phi^{\textrm{K}}_{X_{D_i}}}s\vert h_{{D_i}x}\vert^2 r_{{D_i}x}^{-\alpha_{\textrm{K}}}  \Bigg)}\Bigg]\nonumber \\
&=& \mathbb{E}\Bigg[\prod_{x\in\Phi^{\textrm{K}}_{X_{D_i}}} \exp\Bigg(-s\vert h_{{D_i}x}\vert^2 r_{{D_i}x}^{-\alpha_{\textrm{K}}}\Bigg)\Bigg]\nonumber \\
&\overset{(a)}{=}&\mathbb{E}\Bigg[\prod_{x\in\Phi^{\textrm{K}}_{X_{D_i}}}\mathbb{E}_{\vert  h_{{D_i}x}\vert^2, p}\Bigg\lbrace \exp\Bigg(-s\vert h_{{D_i}x}\vert^2r_{D_ix}^{-\alpha_{\textrm{K}}}\Bigg)\Bigg\rbrace\Bigg]\nonumber \\
&\overset{(b)}{=}&\mathbb{E}\Bigg[\prod_{x\in\Phi^{\textrm{K}}_{X_{D_i}}}\dfrac{p}{1+s r_{{D_i}x}^{-\alpha_{\textrm{K}}}}+1-p\Bigg]\nonumber \\
&\overset{(c)}{=}&\exp\Bigg(-\lambda^{\textrm{K}}_{X}\displaystyle\int_{\mathbb{R}}\Bigg[1-\bigg(\dfrac{p}{1+sr_{{D_i}x}^{-\alpha_{\textrm{K}}}}+1-p\bigg)\Bigg]\textrm{d}x\Bigg)\nonumber \\
&=&\exp\Bigg(-p\lambda^{\textrm{K}}_{X}\displaystyle\int_{\mathbb{R}}\dfrac{1}{1+1/sr_{{D_i}x}^{-\alpha_{\textrm{K}}}}\textrm{d}x\Bigg)\\
&=&\exp\Bigg(-p\lambda^{\textrm{K}}_{X}\displaystyle\int_{\mathbb{R}}\dfrac{1}{1+r_{{D_i}x}^{\alpha_{\textrm{K}}}/s}\textrm{d}x\Bigg), 
\end{eqnarray}
where (a) follows from the independence of the fading coefficients; (b) follows from performing the expectation over $|h_{{D_i}x}|^2$ which follows an exponential distribution with unit mean, and performing the expectation over the set of interferes; (c) follows from the probability generating functional (PGFL) of a PPP. The expression of $\mathcal{L}_{{I^{\textrm{K}}_{Y_{D_i}}}}(s)$ can be acquired by following the same steps. 

 \section{}\label{App_C} 
In order to calculate the Laplace transform of interference originated from the $X$ road at the node $D_i$, we have to calculate the integral in (\ref{eq:33}). We calculate the integral in (\ref{eq:33}) for $\alpha_{\textrm{K}}=2$. We set ${d_i}_{x}=d_i \cos(\theta_{D_i})$, and ${d_i}_{y}=d_i \sin(\theta_{D_i}$), then (\ref{eq:33}) becomes
\begin{equation}\label{eq:65}
\mathcal{L}_{I^{\textrm{K}}_{X_{D_i}}}(s)=\exp\Bigg(-\emph{p}\lambda^{\textrm{K}}_{X}s\int_\mathbb{R}\dfrac{1}{s+{d_i}_{y}^2+(x-{d_i}_{x})^2 }\textrm{d}x\Bigg),
\end{equation}
and the integral inside the exponential in (\ref{eq:65}) equals
\begin{equation}\label{eq:66}
\int_\mathbb{R}\dfrac{1}{s+{d_i}_{y}^2+(x-{d_i}_{x})^2}\textrm{d}x=\dfrac{\pi}{\sqrt{{d_i}_{y}^2+s}}.
\end{equation}
Then, plugging (\ref{eq:66}) into (\ref{eq:65}), and substituting ${d_i}_{y}$ by $d_i\sin(\theta_{{D_i}})$  we obtain
\begin{equation}\label{eq:67}
\mathcal{L}_{I^{\textrm{K}}_{X_{D_i}}}(s)=\exp\Bigg(-\dfrac{\emph{p}\lambda^{\textrm{K}}_{X}s\,\pi}{\sqrt{{d_i}^2\sin(\theta_{{D_i}})^2+s}}\Bigg).
\end{equation}
Following the same steps above, and without details for the derivation with respect to $s$, we obtain 
\begin{equation}\label{eq:68}
\mathcal{L}_{I^{\textrm{K}}_{Y_{D_i}}}(s)=\exp\Bigg(-\dfrac{\emph{p}\lambda^{\textrm{K}}_{Y}s\,\pi}{\sqrt{{d_i}^2\cos(\theta_{{D_i}})^2+s}}\Bigg).
\end{equation}
Then, when compute the derivative of (\ref{eq:67}) and (\ref{eq:68}), we obtain

\begin{multline}\label{eq.40}  
\frac{\textrm{d}^{k-n} \mathcal{L}_{I^{\textrm{K}}_{X_{D_i}}}\big(s\big)}{\textrm{d}^{k-n} s}= \Bigg[-\frac{\emph{p}\lambda^{\textrm{K}}_{X}\pi}{\sqrt{{d_i}^2\sin(\theta_{{D_i}})^2+s}}+\frac{1}{2}\frac{\emph{p}\lambda^{\textrm{K}}_{X}\pi s}{({d_i}^2\sin(\theta_{{D_i}})^2+s)^{3/2}}\Bigg]^{k-n}\\
\times \exp\Bigg(-\frac{\emph{p}\lambda^{\textrm{K}}_{X}\pi s}{\sqrt{{d_i}^2\sin(\theta_{{D_i}})^2+s}}\Bigg).
\end{multline} 
\begin{multline}\label{eq.41} 
  \frac{\textrm{d}^{n} \mathcal{L}_{I^{\textrm{K}}_{Y_{D_i}}}\big(s\big)}{\textrm{d}^{n} s}=\Bigg[-\frac{\emph{p}\lambda^{\textrm{K}}_{Y}\pi}{\sqrt{{d_i}^2\cos(\theta_{{D_i}})^2+s}}+\frac{1}{2}\frac{\emph{p}\lambda^{\textrm{K}}_{Y}\pi s}{({d_i}^2\cos(\theta_{{D_i}})^2+s)^{3/2}}\Bigg]^n \\
\times \exp\Bigg(-\frac{\emph{p}\lambda^{\textrm{K}}_{Y}\pi s}{\sqrt{{d_i}^2\cos(\theta_{{D_i}})^2+s}}\Bigg).
\end{multline} 
\bibliographystyle{ieeetr}
\bibliography{bibnoma}
\end{document}